**Science overlay maps: a new tool for research policy and library management**

**Ismael Rafols[1], Alan L. Porter[2] and Loet Leydesdorff[3]**

[1] SPRU –Science and Technology Policy Research, University of Sussex, Brighton, BN1 9QE, England; i.rafols@sussex.ac.uk ;

[2] Technology Policy and Assessment Center, Georgia Institute of Technology, Atlanta, GA 30332, USA; alan.porter@isye.gatech.edu; and Search Technology, Inc.

[3] Amsterdam School of Communication Research (ASCoR), University of Amsterdam, Kloveniersburg 48, 1012 CX Amsterdam, The Netherlands; loet@leydesdorff.net

**Abstract**

We present a novel approach to visually locate bodies of research within the sciences, both at each moment of time and dynamically. This article describes how this approach fits with other efforts to locally and globally map scientific outputs. We then show how these science overlay maps help benchmark, explore collaborations, and track temporal changes, using examples of universities, corporations, funding agencies, and research topics. We address conditions of application, with their advantages, downsides and limitations. Overlay maps especially help investigate the increasing number of scientific developments and organisations that do not fit within traditional disciplinary categories. We make these tools accessible to help researchers explore the ongoing socio-cognitive transformation of science and technology systems.

**Keywords:** science, map, overlay, classification, interdisciplinary, research, evaluation.

## 1. Introduction

Most science and technology institutions have undergone or are undergoing major reforms in their organisation and in their activities in order to respond to changing intellectual environments and increasing societal demands for relevance. As a result, the traditional structures and practices of science, built around disciplines, are being by-passed by organisation in various ways in order to pursue new types of differentiation that react to diverse pressures (such as service to industry needs, translation to policy goals, openness to public scrutiny, etcetera). However, no clear alternative socio-cognitive structure has yet replaced the "old" disciplinary classification. In this fluid context, in which social structure often no longer matches with the dominant cognitive classification in terms of disciplines, it has become increasingly necessary for institutions to understand and make strategic choices about their positions and directions in moving cognitive spaces. "The ship has to be reconstructed while a storm is raging at sea." (Neurath, 1932/33) The overlay map of science we present here is a technique that intends to be helpful in responding to these needs elaborating on recently developed global maps of science (Leydesdorff & Rafols, 2009).

Although one would expect global maps of science to be highly dependent on the classification of publications, the clustering algorithms, and visualisation techniques used, recent studies comparing maps created using very different methods revealed that, at a coarse level, these maps are surprisingly robust (Klavans & Boyack, 2009; Rafols & Leydesdorff, 2009). This stability allows to 'overlay' publications or references produced by a specific organisation or research field against the background of a stable representation of global science and to produce comparisons that are visually attractive, very readable, and potentially useful for science policy-making or research and library management. In this study, we present one such overlay technique and introduce its possible usages by practitioners by providing some demonstrations. For example, one can assess a portfolio at the global level or animate a diffusion pattern of a new field of research. We illustrate the former application with examples from universities, industries and funding agencies, and the latter for an emergent research topic (carbon nanotubes). In appendices we provide the technical information for making these overlays using software available in the public domain.

Our first objective is to introduce the method for making and/or utilising the global maps to prospective users in the wider science policy and research management communities who are not able to follow the developments in scientometrics in detail. Since the paper addresses a wide audience, we shall not discuss technical bibliometric issues, but provide references to further literature. Secondly, we reflect on issues about the validity and reliability of these maps. Thirdly, this study explores the qualitative conditions of application of the maps, proposing examples of meaningful usage and flagging out potential misreadings and misunderstandings.

As classifications, maps can become embedded into working practices and turn into habit, or be taken for granted away from public debate, yet still shaping policy or management decisions that may benefit some groups at the expense of others (Bowker & Star, 2000, pp. 319-320). In our opinion, scientometric tools remain error-prone representations and fair use can only be defined reflexively. Maps, however, allow for more interpretative flexibility than rankings. By specifying the basis, limits, opportunities and pitfalls of the global and overlay maps of science we try to avoid the widespread problems that have beset the policy and management (mis-)use of bibliometric indicators such as the *impact factor* (Martin, 1997; Gläser & Laudel, 2007). By specifying some of the possible sources of error, we aim to set the conditions so that this novel tool remains open to critical scrutiny and can be used in an appropriate and responsible manner (Rip, 1997, p. 9).

## 2. The dissonance between the epistemic and social structures of science

The traditional representation of science was derived from the so-called 'tree of knowledge' according to which metaphor, knowledge is split into branches, then into major disciplines and further differentiated into subdisciplines and specialties. The modern universities mainly organised their social structure along this model (Lenoir, 1997), with a strong belief that specialisation was key for successful scientific endeavour (Weber, 1919). However, many (if not most) scientific activities no longer align with disciplinary boundaries (Whitley, 1984 (2000); Klein, 2000; Stehr & Weingart, 2000).[1] As Lenoir (1997, p. 53) formulated:

> Scientists at the research front do not perceive their goal as expanding a discipline. Indeed most novel research, particularly in contemporary science, is not confined within the scope of a single discipline, but draws upon work of several disciplines. If asked, most scientists would say that they work on problems. Almost no one thinks of her- or himself as working on a discipline.

The changing social contract of science, progressively enacted in the last 20 years, has brought a stronger focus on socio-economic relevance and accountability (Gibbons *et al*. 1994; Etzkowitz & Leydesdorff, 2000), which has exacerbated the dissonances between epistemic and organisational structures. Descriptions of recent transformations emphasise inter-, multi-, or transdisciplinary research as a key characteristic of the new forms of knowledge production (reviewed by Hessels & Van Lente, 2008).

These ongoing changes pose challenges to the conduct and institutional management of science and higher education. New 'disciplines' that emerged in the last decades, such as computer or cognitive sciences do not fit neatly into the tree of knowledge. Demands for socially relevant research have also led to the creation of mission-oriented institutes and centres targeting societal problems, such as mental health or climate change, that spread (and sometimes cross-fertilise) across disciplines. At the institutional level, however, one cannot avoid the key question of the relative position of these emergent organisations and fields in relation to 'traditional' disciplines when it comes to the evaluation. Can changes in research areas be measured against a baseline (Leydesdorff *et al.*, 1994; Studer & Chubin, 1982)? Are the new developments transient (Gibbons *et al.*, 1994) or, perhaps, just relabeling of "old wine" (Van den Daele *et al.*, 1979; Weingart, 2000)? Such questions point to our endeavour: can science overlay maps be a tool to explore the increasingly fluid and complex dynamics of the sciences? Do they allow us to throw light upon the cognitive and organisational dynamics, thereby facilitating research-related choices (e.g., funding, organization)?

---

[1] The 'tree of knowledge' (e.g., Maturana and Varela, 1984) has strong similarities with the 'tree of life' developed by biology (via the subdicipline of systematics) to explain the diversity of species out of a common origin. Interestingly, the 'tree of life' has also been increasingly challenged by evidence of massive horizontal gene transfer among prokaryotes (Bapteste et al. 2009).

3. **Approaches to mapping the sciences**

Science maps are symbolic representations of scientific fields or organisations in which the elements of the map are associated with topics or themes. Elements are positioned in the map so that other elements with related or similar characteristics are located in their vicinity, while those elements that are dissimilar are positioned at distant locations (Noyons, 2001, p. 84). The elements in the map can be authors, publications, institutes, scientific topics, or instruments, etc. The purpose of the representation is to enable the user to explore relations among the elements.

Science maps were developed in the 1970s (Small 1973; Small & Griffith, 1974; Small & Sweeny, 1985; Small *et al.*, 1985). They underwent a period of development and dispute regarding their validity in the 1980s (Leydesdorff, 1987; Hicks, 1987; Tijssen *et al.*, 1987), and a slow process of uptake in policy during the 1990s, that fell below the expectations created (Noyons, 2001, p. 83). The further development of network analysis during the 1990s made new and more user-friendly visualisation interfaces available. Enhanced availability of data has spread the use and development of science maps during the last decade beyond the scientometrics community, in particular with important contributions by computer scientists specialised in the visualisation of information (Börner et al. 2003), as illustrated by the educative and museological exhibition, *Places and Spaces* (http://www.scimaps.org/ ).

Most science maps use data from bibliographic databases, such as PubMed, Thomson Reuters' Web of Science or Elsevier's Scopus, but they can also be created using other data sources (e.g., course pre-requisite structures, Balaban & Klein, 2006). Maps are built on the basis of a matrix of similarity measures computed from correlation functions among information items present in different elements (e.g. co-occurrence of the same author in various articles). The multidimensional matrices are projected onto two or three dimensions. Details of these methods are provided by Leydesdorff (1987), Small (1999) and reviewed by Noyons (2001, 2004) and Börner *et al.* (2003).

In principle, there are several advantages of using maps rather than relying just on numeric indicators. Maps position units in a network instead of ranking them on a list. As in any data visualisation technique, maps furthermore facilitate the reading of bibliometric information by non-experts—with the downside that they also leave room for manipulating the interpretation of data structures. Second, maps allow for the representation of diverse and large sets of data in a succinct way. Third, precisely because they make it possible to combine different types of data, maps also enable users to explore *different* views on a given issue. This interpretive flexibility induces reflexive awareness about the phenomenon the user is analysing and about the analytical value (and pitfalls) of these tools. Maps convey that bibliometrics cannot provide definite, 'closed' answers to science policy questions (such as "picking the winners"). Instead, maps remain more explicitly heuristic tools to explore and potentially open up plural perspectives in order to inform decisions and evaluations (Roessner, 2000; Stirling, 2008).

While the rhetoric of numbers behind indicators can easily be misunderstood as objectified and normalized descriptions of a reality (the "top-10", etc.), the heuristic, toy-like quality of science maps is self-exemplifying. These considerations are important because '[T[here is a lot of misunderstanding [by users] about the validity and utility of the maps' (Noyons, 2004, p. 238). This is compounded with a current lack of ethnographic or sociological validation of the actual use of bibliometric tools (Woolgar, 1991; Rip, 1997; Gläser & Laudel, 2007).

The vast majority of science maps have aimed at portraying *local* developments in science, using various units of analysis and similarity measures. To cite just a few techniques:

- *co-citations of articles* (e.g. research on collagen, Small, 1977);
- *co-word analysis* (Callon et al., 1986), e.g. translation of cancer research (Cambrosio et al., 2007);
- *co-classification of articles* (e.g. neural network research, Noyons & Van Raan, 1998);
- *co-citations of journals* (e.g. artificial intelligence, Van den Besselaar & Leydesdorff, 1996);
- *co-citation of authors* (e.g. information and library sciences, White & McCain, 1998).

These local maps are very useful to understand the *internal* dynamics of a research field or emergent discipline, but typically they cover only a small area of science. Local maps have the advantage of being potentially accurate in their description of the relations *within* a field studied, but the disadvantage is that the units of analyses and the positional co-ordinates remain specific to each study. As a result, these maps cannot teach us how a new field or institute relates to other scientific areas. Furthermore, comparison among different developments is difficult because of the different methodological choices (thresholds and aggregation levels) used in each map.

Shared units of representation and positional co-ordinates are needed for proper comparisons between maps. In order to arrive at stable positional co-ordinates, a full mapping of science is needed. In summary, two requirements can be formulated as conditions for a global map of science: mapping of a full bibliographic database, and robust classification of the sciences. Both requirements were computationally difficult until the last decade and mired in controversy. The next section explains how these controversies are in the process of being resolved and a consensus on the core structure of science is emerging.

## 4. Global maps of science: the emerging consensus

The vision that a comprehensive bibliographic database contained the structure of science was already present in the seminal contributions of Price (1965). From the 1970's, Henry Small and colleagues at the Institute of Scientific Information (ISI) started efforts to achieve a global map of science. In 1987, the ISI launched the first World Atlas of Science (Garfield, 1987) based on co-citation clustering algorithms. However, the methods used (single-linked clustering) were seen as unstable and problematic (Leydesdorff, 1987). Given the many choices that can be made in terms of units of analysis, measures of similarity/distance, reduction of dimensions and visualisation techniques (Börner et al., 2003), most researchers in the field (including ourselves) expected any global science representations to remain heavily dependent on these methodological choices (Leydesdorff, 2006).

Against these expectations, recent results of a series of global maps suggest that the basic structure of science is surprisingly robust. First, Klavans & Boyack (2009) reported a remarkable degree of agreement in the core structure of twenty maps of science generated by independent groups, in spite of different choices of unit of analysis, similarity measure, classification (or clustering algorithms) or visualisation technique[2].Then, Rafols & Leydesdorff (2009) showed that similar global maps can be obtained using significantly 'dissenting' journal classifications. These validations emphasize bibliometric, rather than expert assessment (Rip, 1997, p. 15), but this seems suitable in considering global science mappings, given that no experts are capable of making reliable judgement on the interrelations of all parts of science (Boyack et al., 2005, p. 359; Moya-Anegón et al, 2007, p. 2172). The consensus is more about the coarse structure of science than on final maps. The latter may show apparent discrepancies due to different choices of representation. This is the case, for example, when one compares Moya-Anegón et al. (2007) use of fully centric maps as opposed to Klavans & Boyack's (2008) fully circular ones.

Let us explore key features of the emerging consensus on the global structure, illustrating with Figure 1. The first feature is that science is not a continuous body, but a fragmentary structure composed of both solid clusters and empty spaces—in geographical metaphors, a rugged landscape of high mountains, and deep valleys or faults rather than plains with rolling hills. This quasi-modular structure (or "near decomposability" in terms of the underlying (sub)systems) can be found at different levels. This multi-level cluster structure is related to the power-law distributions in citations (Katz, 1999). Furthermore, these multi-level discontinuities of science are consistent with qualitative descriptions (Dupré, 1993; Galison & Stump, 1996; Abbot, 2001).

A first view of Figure 1 at the global level reveals a major biomedical research pole (to the left in Figure 1), with molecular biology and biochemistry at its centre, and a major physical sciences pole (to the right in Figure 1), including engineering, physics and material sciences.

---

[2] More recently developed maps also show a high degree of agreement, in spite of using very different methods such as hybrid text/citation clustering (Janssens et al. 2009) or click-stream by users of journal websites (Bollen et al. 2009).

A third pole would be constituted by the social sciences and the humanities (at the bottom left in Figure 1).[3]

The second key feature is that the poles described above are arranged in a somewhat circular shape (Klavans & Boyack, 2009)—rather than a uniform ring, more like an uneven doughnut (a torus-like structure) that thickens and thins at different places of its perimeter. This doughnut shape can best be seen in three-dimensional representations; it is not an artefact produced by the reduction of dimensions or choice of algorithm used for the visualisation. The torus-like structure of science is consistent with a pluralistic understanding of the scientific enterprise (Knorr-Cettina, 1999; Whitley, 2000): in a circular geometry no discipline can dominate by occupying the centre of science; and at the same time, each discipline can be considered as at the centre of its own world.

The torus-like structure explains additionally how the great disciplinary divides are bridged. Moving counter-clockwise from 3 o'clock to 10 o'clock in Figure 1 (see Figure 2 for more details), the biomedical and the physical sciences poles are connected by one bridge that reaches from material sciences to chemistry, and a parallel elongated bridge that stretches from engineering and materials to the earth sciences (geosciences and environmental technologies), then through biological systems (ecology and agriculture) to end in the biomedical cluster. Moving from 10 o'clock to 6 o'clock, one can observe how the social sciences are strongly connected to the biomedical cluster via a bridge made by cognitive science and psychology, and a parallel bridge made by disciplines related to health services (such as occupational health and health policy). Finally, moving from 6 o'clock to 3 o'clock, we observe that the social sciences link back to the physical sciences via the weak interactions in mathematical applications and between business and computer sciences.[4]

---

[3] The social sciences appear as a rather diffuse and small area in these science maps due to lower citation rates. However, a recent study of social sciences on their own shows a cluster as large as the natural sciences (Bollen et al, 2009).

The idea behind the emergent consensus is that the most important relations among disciplines are robust—i.e. they can be elicited in the different maps even when their representations differ in many details of the global science map due to other methodological choices. However, one should not underestimate the differences among maps—particularly since they can illuminate biases. In some cases, the disagreements are mainly visual like those between geographic portrayals (e.g. in Mercator vs. Peters projections): although there are different choices regarding the position and area size of Greenland, they all agree that Greenland lies between North America and west Eurasia. However, in some other cases, disagreements can be significant. For example, the position of mathematics (all math subject categories) in the map remains open to debate. Since different strands of mathematics are linked to different major fields (medicine, engineering, social sciences), these may show as diverse entities in distant positions, rather than as a unitary corpus, depending on classifications and/or clustering algorithms used.

It is important to recognize that the underlying relationships are multidimensional, so various two (and three-) dimensional representations can result. For example, we depicted (in Figure 1) chemistry in the centre and geosciences on the periphery, but a 3D representation would show that the opposite representation is also legitimate. Furthermore, due to reduction of dimensions relative distances among categories need to be interpreted with caution, since two categories may appear to be close without being similar. This is the case, for example, for the categories "paper and wood materials science" and "paleontology" (at the top of our basemap), or "dairy and animal science" and "dentistry" (top left). Categories that are only weakly linked to a few other categories are particularly prone to generate this type of positional 'mirage.' On the other hand, dimensional reduction also means that one can expect 'tunnels,' whereby hidden dimensions closely connect apparently distant spaces in the map. For example, "clinical medicine" and a small subset of engineering are connected via a slim 'tunnel' made by "biomedical engineering and nuclear medicine."

---

[4] Notice that the global map has circular symmetry, some branches develop in parallel over the torus(e.g. geosciences and chemistry). As a result, the creation of a fully uni-dimensional *wheel or circle of science* is very elegant and perhaps very useful, but it involves some distortions beyond the consensus structure (see Klavans and Boyack, forthcoming or http://www.scival.com/).

In summary, the consensus on the structure of science enables us to generate and warrant a stable global template to use as a *basemap*. Several representations of this backbone are possible, legitimate and helpful in bringing to the fore different lights and shadows. By standardizing our mapping with a convenient choice (as shown in Figure 2), we can produce comparisons that are potentially useful for researchers, science managers, or policy-makers. For example, one can assess a portfolio at the global level or animate a diffusion pattern of a new field of research.

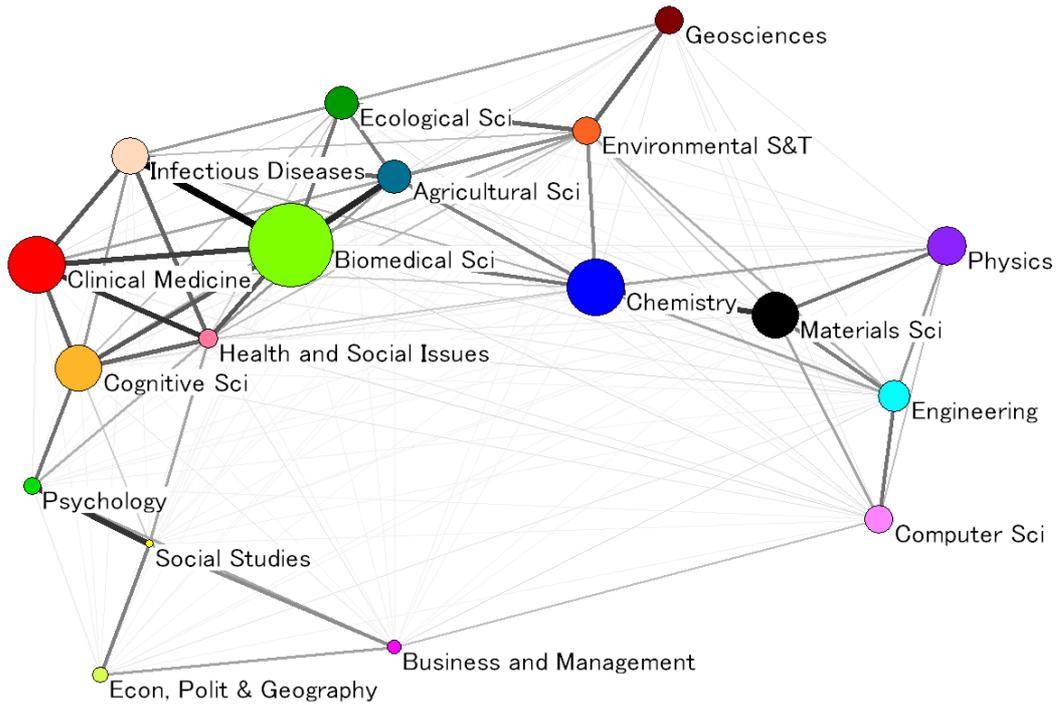

*Figure 1. The core structure of science. Cosine similarity of 18 macro-disciplines created from factor analysis of ISI Subject Categories in 2007. The size of nodes is proportional to number of citations produced.*

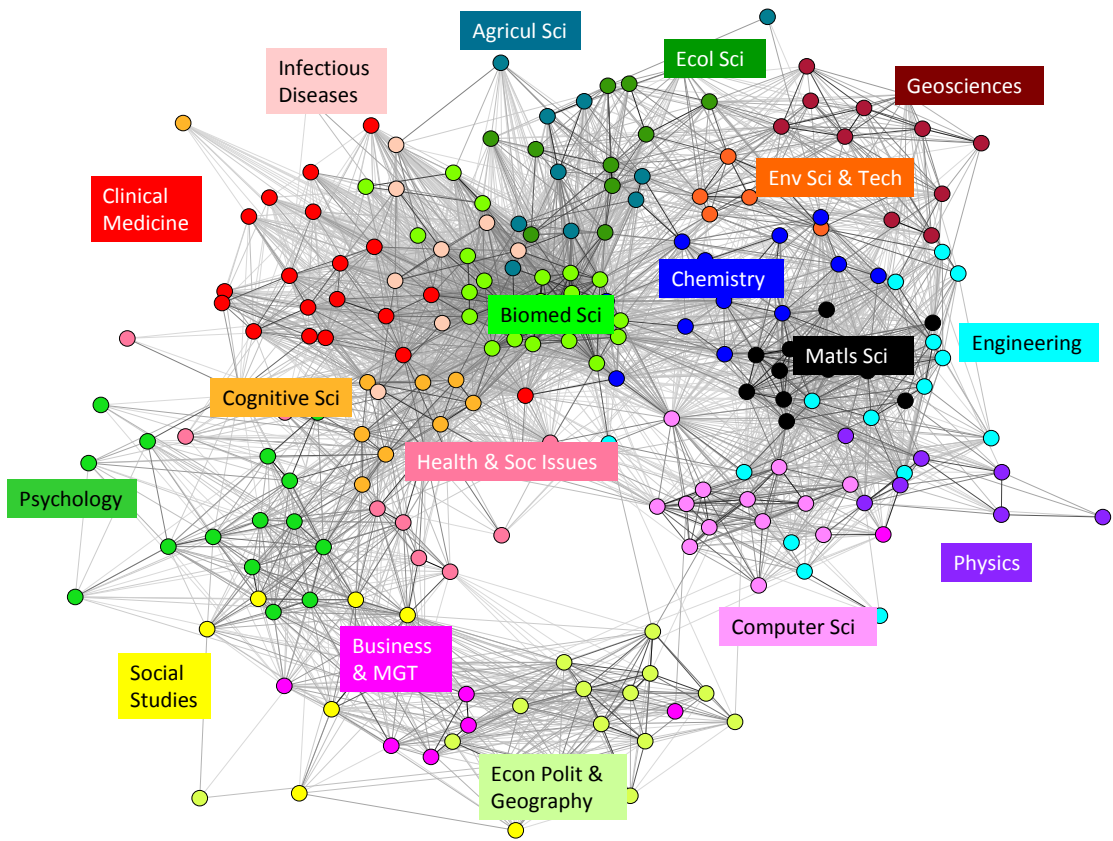

*Figure 2. Global science map based on citing similarities among ISI Subject Categories (2007).*

## 5. Science overlay maps: a novel tool for research analysis

The local science maps are problematic for comparisons because they are not stable in the units or positions of representation, as outlined in section 3. To overcome this, one can use the units and the positions derived from a global map of science, but overlay on them the data corresponding to the organisations or themes under study, as first shown by Boyack (2009). In this section we introduce in detail a method of overlaying maps of science. This method can be explored interactively in our webpage http://idr.gatech.edu/maps or http://www.leydesdorff.net/overlaytoolkit.[5] A step-by-step guide on how to construct overlay maps is provided in Appendix 1.

To construct the basemap, we use the subject categories (SCs) of the Web of Science to which the ISI (Thomson Reuters) assigns journals based on journal-to-journal citation patterns and editorial judgment. The SCs operationalise 'bodies of specialized knowledge' (or subdisciplines) to enable one to track the position of articles. The classification of articles and journals into disciplinary categories is controversial and the accuracy of the ISI classification is open to debate (Pudovkin & Garfield, 2002, at p. 1113n). Other classifications and taxonomies are problematic as well (Rafols & Leydesdorff; 2009; NAS, 2009, p. 22). Bensman & Leydesdorff (2009) argued for using the classification of the Library of Congress, but this extension would lead us beyond the scope of this study. However, since the global maps have been shown to be relatively robust even when there is 50% disagreement about classifications, we pragmatically choose the classification that has been most widely used and is most easily accessible, despite its shortcomings (Rafols & Leydesdorff, 2009; see Appendix 2).

---

[5] A user-friendly toolkit using freeware *Pajek* is available at http://www.leydesdorff.net/overlaytoolkit/sc2007.zip .

We follow the same method outlined in Leydesdorff & Rafols (2009), inspired by Moya-Anegón et al. (2004). First, data were harvested from the CD-Rom version of the Journal Citation Reports (JCR) of the Science Citation Index (SCI) and the Social Science Citations Index (SSCI) of 2007, containing 221 Subject Categories (SCs). This data is used to generate a matrix of citing SCs to cited SCs with a total of 60,947,519 instances of citations between SCs. Salton's cosine was used for normalization in the citing direction. *Pajek* is used for the visualizations (http://pajek.imfm.si) and SPSS (v15) for the factor analysis. Figure 2 shows the global map of science obtained using the 221 ISI SCs in 2007. Each of the nodes in the map shows one SC, representing a subdiscipline. The lines indicate the degree of similarity (with a threshold cutoff at a *cosine* similarity > 0.15) between two SCs, with darker and thicker lines indicating stronger similarity. The relative position of the SCs is determined by the pulls of the lines as a system of strings, depending on the extent of similarity, based on the algorithm of Kamada and Kawai (1989). Although in this case we used the ISI SCs, the same method was reproduced with other classification schemes (Rafols & Leydesdorff, 2009).

The labels and colours in Figure 2 display 18 macro-disciplines (groupings of SCs) obtained using factor analysis of this same matrix. The attribution of SCs to factors is listed in the file *221_SCs_2007_Citations&Similarities.xls* provided in the supplementary materials [6] The choice of 18 factors was set pragmatically since it was found that the 19$^{th}$ factor did not load strongly to its own elements. Figure 1, which we used above to illustrate the discussion on the degree of consensus, shows the core structure of science according to these18 macro-categories.

The full map of science shown in Figure 2 provides the basemap over which we will explore specific organisations or scientific themes using our 'overlay' technique. The method is straightforward. First, the analyst retrieves a set of documents at the Web of Science. This set of documents is the body of research to be studied -- e.g., the publications of an organisation, or the references (knowledge base) used in an emergent field, or the citations (audience) to the publications of a successful laboratory. By assigning each document to a category, the function *Analyze* provided in the Web of Science interface can be used to generate a list of the number of documents present in each SC. Uploading this list, the visualization freeware *Pajek* produces a map of science in which the size of a node (SC) is proportional to the number of documents in that category. Full details of the procedure to generate this vector are provided in Appendix 1.

---

[6] This matrix is also available at http://www.leydesdorff.net/overlaytoolkit/SC2007.xls .

Figure 3 illustrates the use of science overlay maps by comparing the profiles of three universities with distinct strengths: the University of Amsterdam, the Georgia Institute of Technology, and the London School of Economics (LSE). For each of them, the publications from 2000 to 2009 were harvested and classified into SCs in the Web of Science.[7] The maps show that the University of Amsterdam is an organisation with a diverse portfolio and extensive research activity in clinical medicine. Georgia Tech is strong in computer sciences, materials sciences, and engineering—as well as in applications of engineering, such as biomedical or environmental technologies. Not surprisingly, LSE's main activity lies in the areas of 1) politics, economics and geography, and 2) social studies—with some activity in the engineering and computer sciences with social applications (e.g. statistics, information systems, or operations research) and in the health services (e.g. heath care and public health). To fully appreciate the descriptions, labels for each of the nodes are needed. Although they are not presented in these figures due to lack of resolution in printed material, labels can be switched on and off in the computer visualisation interface, as explained in Appendix 1.

---

[7] For the University of Amsterdam 31,507 publications were retrieved; 26,701 for Georgia Tech; and 6,555 for LSE.

Some of the advantages of overlay maps over local maps are illustrated by Figure 3. First, they provide a visual framework that enables us to make immediate and intuitively rich comparisons. Second, they use cognitive units for the representation (disciplines and specialties) that fit with conventional wisdom, whereas one can expect the analytical aggregates of local maps to be unstable and difficult to interpret. Third, whereas the generation of meaningful local maps requires bibliometric expertise, overlay maps can be produced by users of the *Science Citation Index* who are not experts in scientometrics. Finally, they can be used for various purposes depending on the units of analysis displayed by the size of the nodes: whether number of publications, citing articles, cited references, growth or other indicators as shown by a series of recent studies (cf. Rafols & Meyer, forthcoming; Porter & Rafols, 2009; Porter & Youtie, 2009).

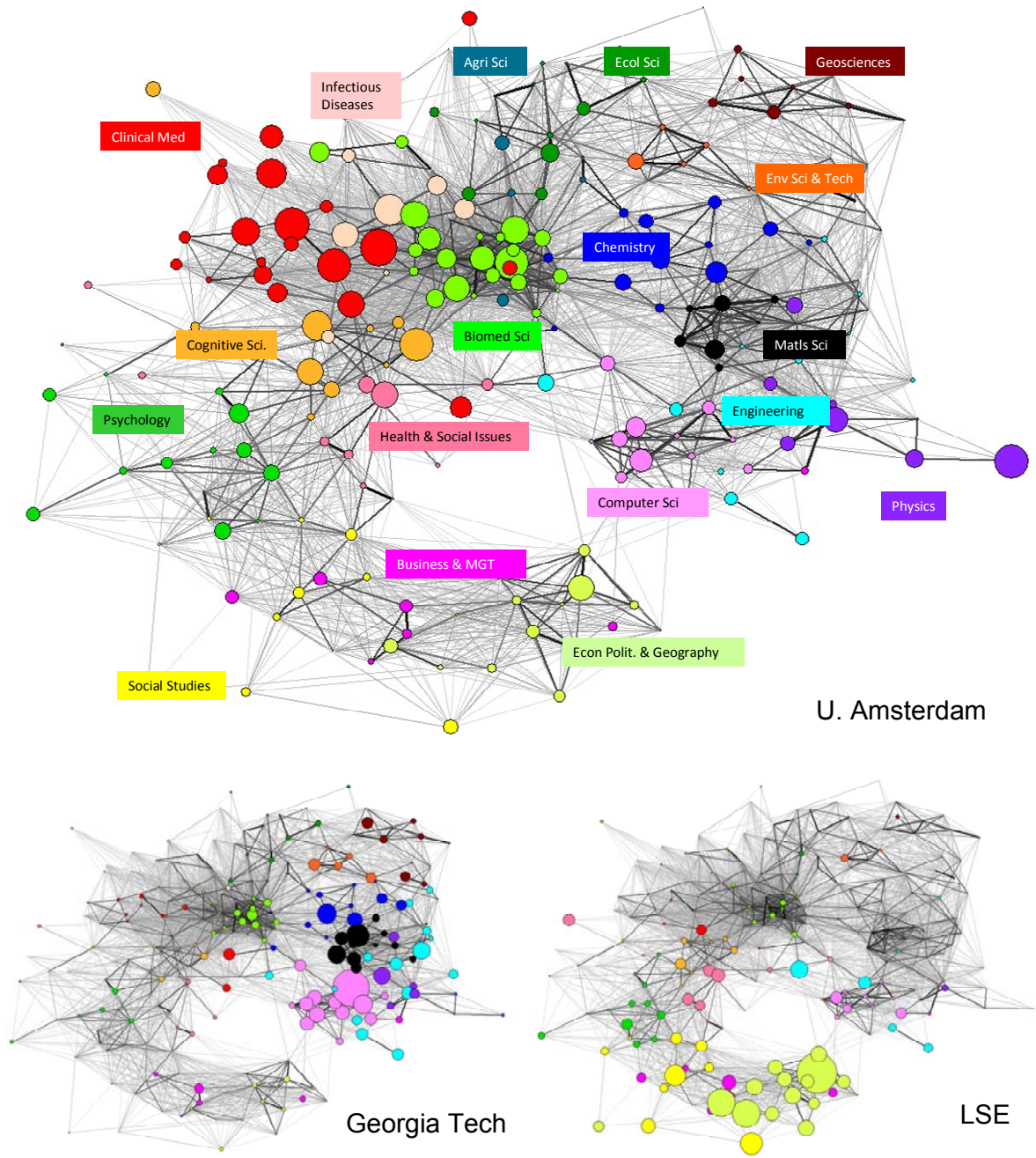

*Figure 3. Publications profiles of the University of Amsterdam, Georgia Tech and London School of Economics (LSE) overlaid on the map of science.*

## 6. Conditions of application of the overlay maps

As is the case with all bibliometric indicators, the appropriate use of overlay maps should not be taken for granted, particularly since they are tools that can be easily used by non-experts (Gläser & Laudel, 2007). In this section we explore the conditions under which overlay maps can be valid and useful for science policy analysis and management, building on Rip (1997).

A first issue concerns the use of journals as the basic unit for classification. This is inaccurate since journals can be expected to combine different epistemic foci, and scientists can be expected to read sections and specific articles from different journals (Bensman 2007). Furthermore, journal content may not fully match specific categories. In particular, consider journals such as *Nature* and *Science* that cover multiple fields. The ISI includes these in their category "Multidisciplinary Sciences" (which is factored into our Biomedical Sciences macro-discipline, even though physics, chemistry, etc., articles appear in it). To date, we just treat this and the seven other interdisciplinary or multidisciplinary SCs (e.g., "Chemistry, Multidisciplinary") the same as any other SCs.

However, the structural similarity of maps obtained with different classifications suggests that discrepancies and errors are not biased and therefore tend to average out when aggregated (Rafols & Leydesdorff, 2009). Hence, the answer to the problem of generalizing from specific or local data to a global map lies in the power of statistics: given a sufficiently large number of assignations, there is high probability that the largest number of publications will have been assigned correctly. Assuming a category with an expected correct assignment of 50%, the binomial test predicts that about 70 papers are sufficient to guarantee the correct assignation of at least 40% of the papers to this category with a significance level of 0.05. Appendix 2 provides further details of the binomial test and estimates of the minimum size of samples under different constraints.[8] These results suggest that one should be cautious about asserting how accurately we are "locating" a given body of research based on small numbers of papers. Instead, for the study of single researchers or laboratories, it may be best to rely on proxies. For example, if a researcher has 30 publications, the analyst is advised to consider the set of references within these articles as a proxy for the disciplinary profile (Rafols & Meyer, forthcoming).

A second set of conditions for the overlay maps to be useful for research policy and management purposes is transparency and traceability, i.e., being able to specify, reproduce, and justify the procedures behind the maps in the public domain. Although the majority of the users of the map may not be interested in the scientometric details, the possibility to re-trace the methods and challenge assumptions is crucial for the maps to contribute to policy debates, where transparency is a requirement. For example, Rip (1997) noted that in the politically charged dispute regarding the 'decline' of British science in the 1980s, a key issue of debate concerned the use of static versus dynamic journal categories (Irvine et al. 1985; Leydesdorff, 1988).

A further requirement for traceability, is relative parsimony, that is, the rule to avoid unnecessary complexity in procedures and algorithms so that acceptable representations can be obtained by counter-expertise or even non-experts—even at the expense of some detailed accuracy—in order to facilitate public discussion, if needed be. In the case of overlay maps, traceability involves making publicly available the following choices: the underlying classifications used and/or clustering algorithms to obtain them (in our case, the ISI SC's); the similarity measures used among categories (Salton's cosine similarity); and the visualisation techniques (Kamada-Kawai with a cosine > 0.15 threshold). These minimal requirements are needed so that the maps can be reproduced and validated independently.

---

[8] This result of at least 70 papers for each of the top categories to be identified in an overlay map is obtained under a very conservative estimate of the accuracy of existing classifications—less stringent estimates suggest that some 10-20 papers per top category may provide overlay maps with accuracy within the vicinity of a SC.

A third condition of application concerns the appropriateness of the given science overlay map for the evaluation or foresight questions that are to be answered. Roessner's (2000) critique of the indiscriminate use of quantitative indicators in research evaluation applies also to maps: without a clear formulation of the question of what a programme or an organisation aims to accomplish, and its context, science maps cannot provide a well-targeted answer. What type of questions can our overlay maps help to answer? We think that they can be particularly helpful for comparative purposes in benchmarking collaborative activities and looking at temporal change, as described in the next section.

## 7. Use in science policy and research management

The changes that S&T systems are undergoing exacerbate the apparent dissonance between social and cognitive structures—with new cross-disciplinary or transversal co-ordinates (Whitley, 2000, p. xl ; Shinn & Ragouet, 2005). As a result, disciplinary labels of university or R&D units cannot be relied upon to provide an accurate description of their epistemic activities. This is because researchers often publish outside the field of their departmental affiliation (Bourke & Butler 1998) and, further, cite outside their field of publication (Van Leeuwen & Tijssen, 2000)—and increasingly so (Porter & Rafols, 2009).

Science overlay maps offer a method to locate or compare positions, shifts and/or dissonances in the disciplinary activities at different institutional or thematic levels. This type of map (with a different basemap) was first introduced by Kevin Boyack and collaborators to compare the disciplinary differences in the scientific strength of nations,[9] in the publishing profiles of two large research organisations (Boyack, 2009, pp. 36-37), and the publication

---
[9] http://wbpaley.com/brad/mapOfScience/index.html; accessed December 15, 2009.

outcomes of two funding agencies (Boyack, Börner & Klavans, 2009, p. 49). Some of us have used previous versions of the current overlay method to

- compare the degree of interdisciplinarity at the laboratory level (Rafols & Meyer, forthcoming);
- study the diffusion of a research topic across disciplines (Kiss et al., 2009);
- model the evolution over time of cross-disciplinary citations in six established research fields (SCs -- Porter & Rafols, 2009);
- explore the multidisciplinary knowledge bases of emerging technologies, namely nanotechnology, as a field (Porter & Youtie, 2009) and specific sub-specialties (Rafols et al., 2010; Huang *et al.*, forthcoming).

The following examples focus on applications for the purposes of benchmarking, establishing collaboration and capturing temporal change, as illustrated with universities (Figure 3), large corporations (Figure 4), funding agencies (Figure 5), and an emergent topic of research (carbon nanotubes, Figure 6).[10]

**Benchmarking**

A first potential use of organisational comparisons is benchmarking: how is organisation A performing in comparison to possible competitors or collaborators? For example a comparison between Pfizer (Figure 4) and Astrazeneca (not shown), reveals at first glance a very similar profile, centred around biomedical research (pharmacology, biochemistry, toxicology, oncology) with activity both in clinical medicine and chemistry. However, a more careful look allows spotting some differences: whereas Pfizer has a strong profile in nephrology, Astrazeneca is more active in gastroenterology and cardiovascular systems.

---

[10] The data shown were retrieved from Web of Science in October, 2009. Figure 4 is based on 8107 publications by Pfizer, 1772 by Nestlé, 2632 by Unilever, and 1617 by Shell between 2000 and 2009. Figure 5 is based on 42,440 publications funded by NIH, 40,283 by NSF, 2,104 by BBSRC, and 5,746 by EPSRC, using the new field of "funding agency." Figure 6 is based on 7,782 publications on carbon-nanotubes in 2008 (cf. Lucio-Arias and Leydesdorff, 2007).

This description may be too coarse for some purposes (e.g. specific R&D investment), but sufficient for policy-oriented analysts to discuss the knowledge base of the firms.[11]

Several choices can be made regarding the data to be displayed in the maps. First, should the map display an input (the categories of the *papers cited by* the organisations), an output (the categories of a set of *publications* of the organisation), or an outcome (the categories of the *papers citing* the organisation's research)? Second, should the overlay data be normalised by the size of the category and/or the size of the organisation? [The figures here are normalised by the size of the organisation, but not by the size of the category; normalising by category will bring to the forefront those categories in which one organisation is relatively very active compared to others, even if it represents a small percentage of its production.] Third, in addition to the number or proportion of publications per SC (or macro-discipline), other indicators such as impact factor or growth rate indicators can be mapped (Noyons et al., 1999; Van Raan & Van Leeuwen, 2002; or Klavans & Boyack, forthcoming).

---

[11] Personal communication with an analyst in a pharmaceutical corporation, November 2009.

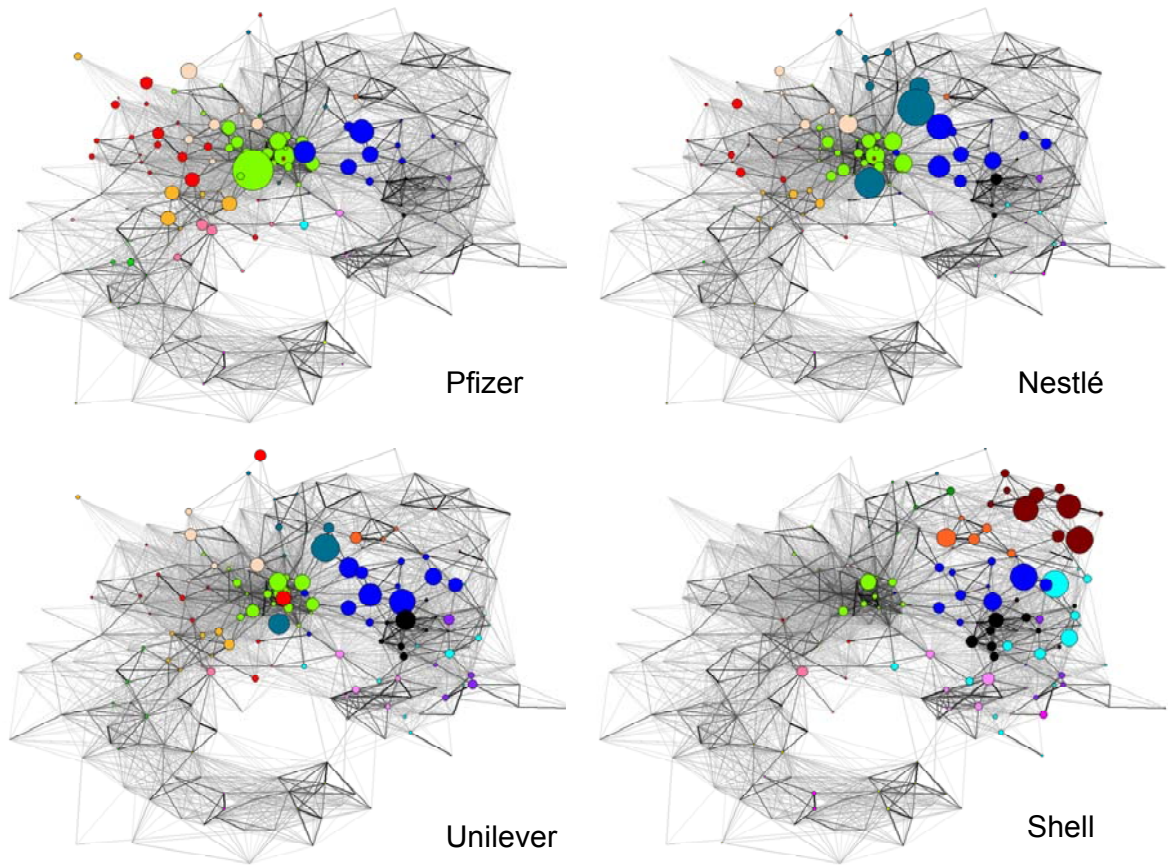

*Figure 4. Profiles of the publications (2000-2009) of large corporations of different economic sectors: pharmaceutical (Pfizer), food (Nestlé), consumer products (Unilever), oil (Shell).*

**Exploring collaborations**

A second application of the overlay maps is to explore complementarities and possible collaborations (Boyack, 2009). For example, Nestlé's core activities lie in food-related science and technology. Interestingly, the map reveals that one of its areas of highest research publication activity, the field of nutrition and dietetics (the dark green spot in the light green cluster in Figure 4 for Nestlé), falls much closer to the biomedical sciences than other food-related research. This suggests that the field of nutrition may act as bridge and common ground for research collaboration between the food and pharmaceutical industry—sectors that are approaching one another, as shown by Nestlé's strategic R&D investment in 'functional' (i.e. health-enhancing) foods (*The Economist*, 2009).

In Figure 5, we compare funding agencies in terms of potential overlap. The funding agencies in the US and the UK have, in principle, quite differentiated remits. In the US (top of Figure 5), the NIH (National Institute of Health) focuses on biomedical research while the NSF (National Science Foundation) covers all basic research. In the UK (bottom of Figure 5), the BBSRC (Biotechnology and Biological Sciences Research Council) and the EPSRC (Engineering and Physical Sciences Research Council) are expected to cover the areas described in their respective names. However, Figure 5 reveals substantial areas of overlap. These are areas where duplication of efforts could be occurring—suggesting a case for coordination among agencies. It may also help indentify interdisciplinary topics warranting express collaboration between committees from two agencies, such as the interaction of the BBSRC and EPSRC on Engineering and Biological Systems.

The exploration of collaboration practices is a topic where overlay maps provide added value because they implicitly convey information regarding the cognitive distance among the potential collaborators. A variety of studies (Llerena & Meyer-Krahmer, 2004; Cummings & Kiesler, 2005; Noteboom et al., 2007; Rafols, 2007) have suggested that successful collaborations tend to occur in a middle range of cognitive distance, whereupon the collaborators can succeed at exchanging or sharing complementary knowledge or capabilities, while still being able to understand and coordinate with one another. At short cognitive distances, the benefits of collaboration may be too low to be worth the effort (or competition may be too strong), while at large distances, understanding between partners may become difficult. It remains an empirical question whether one may think of an 'optimal cognitive distance' which would allow formulating a research project with 'optimal diversity' (Van den Bergh, 2008).

In any case, overlay maps offer a first (yet crude) method to explore complementarities between prospective partners. US managers of grant programmes for highly innovative research pointed out to us that the science overlay maps might be useful for finding partners, as well as for evaluating prospective grantees. The U.S. National Academies Keck Futures Initiative (NAKFI) has found it helpful to overlay research publications pertaining to a prospective workshop topic (synthetic biology) to help identify research communities to include.

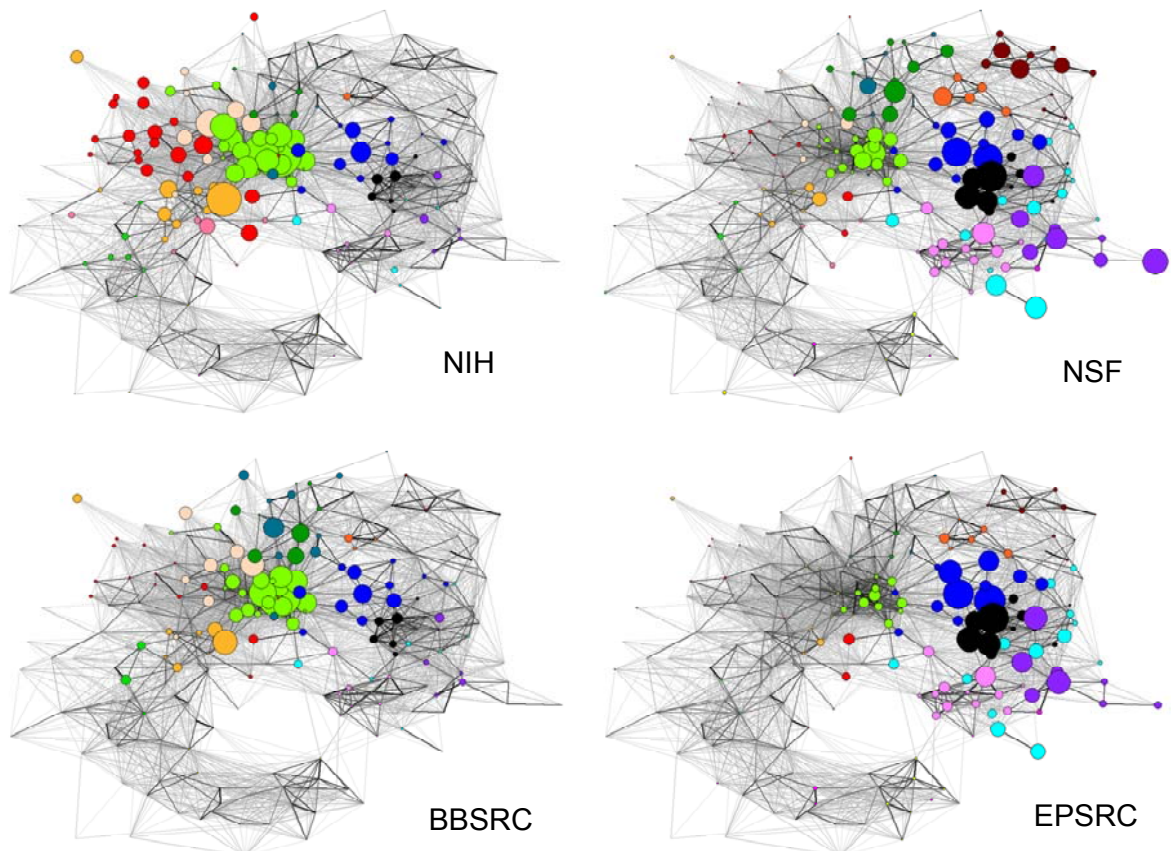

*Figure 5. Publication profile of funding agencies in the US (top) and the UK (bottom).*

**Capturing temporal change**

A third use of overlap maps is to compare developments over time. This allows exploring the diffusion of research topics across disciplines (Kiss et al., 2009). In cases where the research topic is an instrument, a technique or a research material, the spread may cover large areas of the science map (as noted by Price, 1984, p. 16). Figure 6 shows the location of publications on carbon-nanotubes (left) and its areas of growth (right). The growth rate was computed by calculating the annual growth between 2004 and 2008 and taking the average over the period. Since their discovery in 1991, carbon-nanotubes research has shown exponential growth, first in the areas of materials sciences and physical chemistry (Lucio-Arias & Leydesdorff, 2007). However, nowadays the highest growth can be observed in computer sciences due to electronic properties of carbon-nanotubes (pink), in medical applications (red: e.g., imaging and biomedical engineering), and both in biomedical research (green: e.g. pharmacology and oncology) and in environmental research (orange). Within the dominant areas of chemistry and materials sciences (blue and black), growth is highest in applied fields, such as materials for textiles and biomaterials. The overlay methodology thus offers a perspective of the shift of carbon-nanotubes research towards applications and issues of health and environmental safety. Alternatively to a static display of growth rate, the overlay maps can make a "movie" of the evolution of a field (e.g., via a succession of Powerpoint time-slice slides; this works because of the stable basemap).

Comparison over time can also be interesting in order to track developments in organisations. For example, Georgia Tech, traditionally an engineering-centred university without a medical school, recently created the School of BioMedical Engineering. Going back to Figure 3, we can see a medium-size red spot in Georgia Tech publications corresponding to biomedical engineering. A dynamic analysis would depict how this has grown in the last decade.

Since the rationales of research policy, evaluation and management are more complex than bibliometric indicators or maps can be, science overlay maps will usually provide complementary inputs to support (and sometimes to justify) decisions. Other possible uses include finding reviewers for the assessment of interdisciplinary research in emergent fields, or finding valid benchmarks when comparing organisations (Laudel & Gläser, 2009).

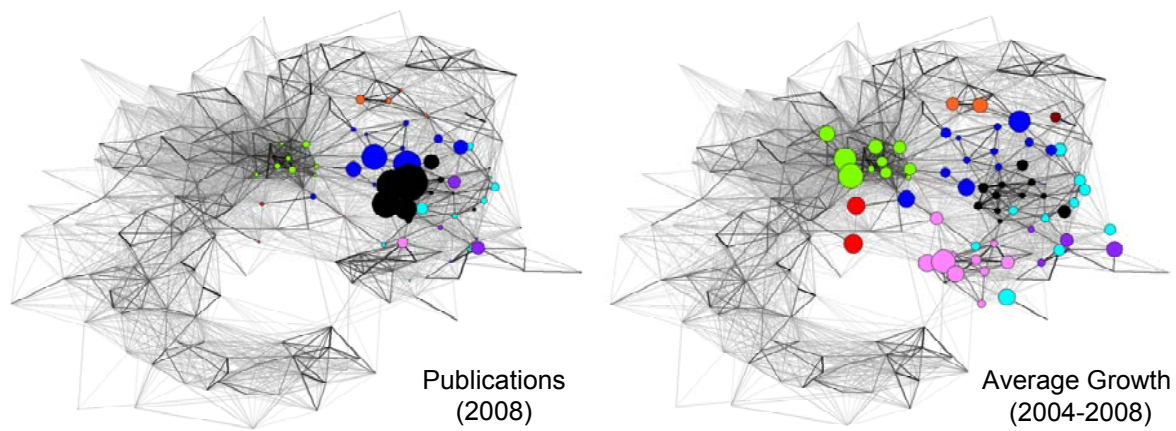

*Figure 6. Publications (2008) and average annual growth of publications (2004-2008) on carbon nanotubes.*

## 8. Advantages and limitations of overlays

We noted above as some major advantages and downsides of overlay maps: on the plus side, their readability, intuitive and heuristic nature; on the minus side, the inaccuracy in the attribution to categories and the possible error by visual inspection of cognitive distance given the reduction of dimensions. In this section, we explore further potential benefits of maps in terms of cognitive contextualisation and capturing diversity, and its main limitation, namely its lack of local relational structures.

**Contextualising categories**

Science overlay maps provide a concise way to contextualise previously existing information of an organisation or topic, in a cognitive space. The same information overlaid on the maps may well have been provided in many previous studies in tabular or bar chart format. For example, policy reports (e.g., Van Raan & Van Leeuwen, 2002) may extensively show the outcomes of a research programme via tables and bar charts: fields of publication, user fields, relative impacts, changes of these indicators over time, etc. What would the overlay maps offer more than this? In our opinion, these maps provide the contextualisation of the data. This extension not only facilitates the comprehension of sets of data, but also their correct interpretation. Unlike bar charts and tables based on categories, the overlay maps remain valid (statistically acceptable) despite possible errors in the classifications. The reason is that, whereas different classifications may produce notably different bar charts, in corresponding maps 'misclassified' articles fall in nearby nodes and the user may still be provided with an adequate pattern. The context can thus reduce perceptual error.

For example, let us consider the new ISI SC of *Nanoscience and Nanotechnology.* A study of a university department in materials science during the 2000s might suggest a strong shift towards nanotechnology based on considering bar charts that show its strong growth in this new SC. However, on a global map of science, this new SC, *Nanoscience and Nanotechnology,* locates extremely closely to other core disciplines in Materials Sciences. Therefore, one would appreciate this change as a relatively small shift in focus, rather than a major cognitive shift. If a department under study had fully ventured into more interdisciplinary nanotechnology, its publications would also increasingly be visible in more disparate disciplines, for example, in the biomedical or environmental areas (Porter & Youtie, 2009).

**Capturing diversity**

Science overlay maps provide the user with a perspective of the disciplinary diversity of any given output, yet without the need to rely on combined or composite indices. Research organisations often seek a diverse cognitive portfolio, but find it difficult to assess whether the intended diversity is achieved. However, diversity encapsulates three entangled aspects (variety, balance, and disparity) which cannot be univocally subsumed under a single index (Stirling, 2007), but are differently reflected in these maps:

- First, the maps capture the *variety* of disciplines by portraying the number of disciplines (nodes) in which a research organisation is engaged;
- Second, they capture the disciplinary *balance* by plotting the different sizes of the SC nodes;

- Third—different from, say, bar charts—maps can convey the *disparity* (i.e. the cognitive distances) among disciplines by placing these units closer or more distant on the map (Rafols & Meyer, forthcoming).

This spatial elaboration of diversity measures is particularly important when comparing scientific fields in terms of multi- or interdisciplinarity. For example, Porter & Rafols (2009) show that in fields such as biotechnology, many disciplines are cited (high variety, a mean of 12.7 subject categories cited per article in 2005), but they are mainly cited in the highly dense area around biomedical sciences (low disparity). In contrast, atomic physics publications cite fewer disciplines (a mean of 8.7 per article), but from a more diverse cognitive area, ranging from physics to materials science and chemistry (higher disparity).

This discussion highlights that overlay maps are useful to explore interdisciplinary developments. In addition to capturing disciplinary diversity, they can also help to clarify the relative location of disciplines and thereby enable us to gain insights of another of the aspects of interdisciplinary research, namely their position *in between* or central (or marginal) to other research areas (Leydesdorff, 2007). Unlike indicators that seek to digest multiple facets to a single value or ranking of the extent of "interdisciplinarity," maps invite the analyst to more reflexive explorations and provide a set of perspectives that can help to open the debate. This plurality is highly commendable given the conspicuous lack of consensus on the assessment of interdisciplinarity (Rinia *et al.* 2001, Morillo *et al.*, 2003; Bordons *et al.*, 2004; Leydesdorff, 2007; Porter *et al.* 2007; Rafols & Meyer, forthcoming; see review by Wagner *et al.*, submitted).

**Missing the relational structure**

The two characteristics that make overlay maps so useful for comparisons, their fixed positional and cognitive categories, are also inevitably, their major limitations and a possible source of misreading. Since the position in the map is only given by the attribution in the disciplinary classification, the resulting map does not teach us anything about the direct linkages between the nodes. For example, Figure 3 shows that the University of Amsterdam covers many disciplines—but we do not know at all whether its local dynamics is organised within the disciplines portrayed or according to a variety of themes transversal to a collection of SCs. In order to investigate this, one would need to create local maps, as described in Section 3. For most local purposes these maps will be based on smaller units of analysis, such as words, publications or journals, rather than SCs.

In our opinion, a particularly helpful option is to combine overlay maps (based on a top-down approach, with fixed and given categories) with local maps (based on a bottom-up approach, with emergent structures), in order to capture the dynamics of an evolving field (Rafols & Meyer, forthcoming; Rafols et al., 2010; Rosvall & Bergstrom, 2009). A recursive combination of overlay and local maps allows us to investigate the evolution of a field both in terms of its *internal* cognitive coherence and the diversity of its knowledge sources with reference to disciplinary classifications (*external*).

## 9. Conclusions

Science overlay maps offer a straightforward and intuitive way of visualising the position of organisations or topics in a fixed map based on conventional disciplinary categories. By thus standardizing the mapping, one can produce comparisons which are easy to grasp for science managers or policy-makers. For example, one can assess a research portfolio of a university or animate a diffusion pattern of an emergent field.

In this study, we have introduced the bases for the use of overlay maps to prospective non-expert users and described how to create them. We demonstrated that the emergent consensus on the structure of science enables us to generate and warrant a stable global template to use as a *basemap.* We introduced the conditions to be met for a proper use of the maps, including a sample size of statistical reliability, and the requirements of transparency and traceability. We provided examples of benchmarking, search of collaborations and examination of temporal change in applications to universities, corporations, funding agencies and emergent topics.

In our opinion, overlay maps provide significant advantages in the readability and contextualisation of disciplinary data and in the interpretation of cognitive diversity. As it is the case with maps in general, overlays are more helpful than indicators to accommodate reflexive scrutiny and plural perspectives. Given the potential benefits of using overlay maps for research policy, we provide the reader with an interactive webpage to explore overlays (http://idr.gatech.edu/maps) and a freeware-based toolkit (available at http://www.leydesdorff.net/overlaytoolkit ).

**Acknowledgements**

ALP and IR acknowledge support from the US National Science Foundation (Award #0830207, 'Measuring and Tracking Research Knowledge Integration', http://idr.gatech.edu/ ). The findings and observations contained in this paper are those of the authors and do not necessarily reflect the views of the National Science Foundation. IR is partially supported by the EU FP7 project FRIDA, grant 225546. We are indebted to K. Boyack, R. Klavans, F. de Moya-Anegón, B. Vargas-Quesada and M. Zitt for insights and discussions.

**Appendix 1: A user-friendly method for the generation of overlay maps**

We follow the method introduced in Rafols & Meyer (forthcoming) to create the overlay map on the basis of a global map of science (Leydesdorff & Rafols, 2009). The steps described below rely on access to the Web of Science and the files available in our mapping kit (http://www.leydesdorff.net/overlaytoolkit). The objective is to obtain the set of SCs for a given set of articles; provide this to network software (we describe for Pajek); and output as overlay information to add to a suitable basemap.

First, the analyst has to conduct a search in the Thomson Reuters Web of Science (www.isiknowledge.com). Non-expert users should note that this initial step is crucial and should be done carefully: authors may come with different initials, addresses are often inaccurate, and only some types of document ,may be of interest (e.g., only so-called citable items: *articles*, *proceedings papers*, *reviews*, and *letters*). Once the analyst has chosen a set of documents from searches at Web of Science, one can click the tab, *Analyze results*. In this new webpage, the selected document set can then be analysed along various criteria (top left hand tab). The *Subject Area* choice produces a list with the number of documents in each Subject Category. This list can be downloaded as *Analyze.txt*. In the next step the analyst can go to our webpage for maps (http://idr.gatech.edu/maps ) and upload this file .

If one analyst desires more control on the process, she can use the programme Pajek and the associated overlaytoolkit. After opening Pajek, press F1 and upload the basemap file *SC2007-015cut-2D-KK.paj*. This files provide the basemap, [12] as shown by selecting *Draw>Draw-Partition-Vector* (or pressing *Ctrl-P*). Then the previously downloaded *Analyze.txt* file has to be transformed by the mini-programme *SC2007.exe* (in our tool kit) as into the Pajek vector format "*SC07.vec*" This file can be uploded into Pajek by choosing

---

[12] The matrix underlying the basemap and the grouping of SCs is available at:
http://www.leydesdorff.net/overlaytoolkit/sc2007.xls

*File>Vector>Read* from the main Pajek menu. Selecting from the menu *Draw>Draw-Partition-Vector* (alternatively, pressing *Ctrl-Q*), the overlay map will be generated. At this stage, the size of nodes will often need adjustment, which can be done by selecting *Options>Sizeof Vertices* in the new draw window. In order to have the *standard* colour settings, the file *SC2007-18Factors-ColourSettings.ini* can be loaded by going to *Options>Ini File>Load* in the main Pajek window. *Crtl-L* and *Ctrl-D* allow visualise and delete, respectively, the labels of each SC. Clickling on nodes allows to move SC to other positions. The image can be exported selecting *Export>2D>Bitmap* in the menu of the *Draw* window. A further optional step would be to label the map in terms of factors, by importing this image into powerpoint in order to label groups of clusters, as shown in the file *SC2007 Global maps.ppt*.

An alternative procedure for more experienced users is to download the records of a document set found in the Web of Science. This is done by adding the *Marked list* (bottom bar) the desired documents; second, going to *Marked list* (top bar) and then downloading the documents in a *Tagged Field* format after selecting *Subject category* as one of the fields to download. The downloaded file should be renamed as *data.txt* and used as input into the program ISI.exe (available at http://www.leydesdorff.net/software/isi). One of the outputs of the programmes ISI.exe is the file *SC07.vec* that can be used in Pajek as explained above. The advantage of this procedure is that ISI.exe also produces other files with information on fields such as author or journal that may be of interest. Feel free to contact the authors in case of difficulty.

**Appendix 2: Estimation of number of papers needed for reliability in overlay maps**

In a previous study, Rafols & Leydesdorff (2009) found that there is between 40-60% of disagreements between attributions of journals to disciplinary categories. Taking a conservative approach, let us assume that for a sample of *N* papers, there is a probability *p=0.5* that they will be misclassified by a given classification (whatever the one that is used). How large should a sample of papers be so that, in spite of the error, the largest categories in the distribution correctly represent the core discipline of the population?

Let us then assume that we have N papers of one given category A. Given the *p=0.5* probability of correct assignation, we only expect 50% of the papers in category A. The analyst has then to arbitrarily choose a lower threshold *m* (we suggest 40%), as the minimum percentage acceptable, with a given degree of significance (we suggest σ=0.05, corresponding to a *z*-score of 1.65). Since a given paper can either be correctly or incorrectly assigned to a category, we can use the binomial distribution to make a binomial test. For *N* ≥ 50 and *Np(1 – p)* ≥ 9, the binomial distribution can be approximated to the normal distribution, with the following z-score:

$$z = \frac{N(p-m)}{\sqrt{Np(1-p)}}$$

For a given degree of significance σ, the associated *z*-score allows us to calculate then the minimum size of the population *N* to guarantee that the correct category A will have at least a proportion *m* in the sample.

$$N \geq \left(\frac{z_{critical}}{p-m}\right)^2 p(1-p)$$

Assuming a degree of mis-classification of 50%, enforcing a lower acceptance threshold of 40% and a significance level of $\sigma=0.05$, one obtains that the sample should be larger than approximately 70 papers (140 papers would increase the significance level to $\sigma=0.01$). Table 1 shows the number of papers needed under different assumptions.

*Table 1. Approximate number of papers recommended for the reliable identification of a category in an overlay map*

| *p* (Prob. disagreement) | *m* (Lower tolerance) | $\sigma$ (Significance level) | *Minimuml number of papers needed* |
|---|---|---|---|
| 0.5 | 0.4 | 0.10 | 41 |
| 0.5 | 0.4 | 0.05 | 68 |
| 0.5 | 0.4 | 0.01 | 135 |
| 0.6 | 0.5 | 0.05 | 65 |
| 0.4 | 0.3 | 0.05 | 65 |

These results teach us the number of papers *N* needed in the populated categories to have some certainty. This means that the actual number of papers per overlay map depends on how narrow or wide is the distribution of disciplinary categories. The more skewed the distribution, the fewer papers are needed. Taking the example of Figure 3, one can estimate that in diverse universities such as Amsterdam or Sussex 3,000 publications may be needed to capture precisely the five top disciplines, whereas for focused organisations such as the European Molecular Biology Laboratory (EMBL), 1,500 publications could be enough.

In our opinion, these are rather conservative estimates, having set at *p=0.5* of mis-assignment. If one allows for 'near-misses' (i.e. assignment to the two nearest categories to be counted as correct) then *p* can be estimated in the range of *0.70* to *0.85* (Rafols & Leydesdorff, 2009). In this case only some dozens of papers are needed to achieve *m~0.5.* (but in this case, the normal distribution approximation cannot be used for the estimate).